# A tentative model for estimating the compressibility of rock-salt $AgCl_xBr_{1-x}$ alloys

VASSILIKI KATSIKA-TSIGOURAKOU[*] and EFTHIMIOS S. SKORDAS

*Department of Solid State Physics, Faculty of Physics, University of Athens, Panepistimiopolis, 157 84 Zografos, Greece*

**Abstract**

Ab initio detailed calculations of the elastic properties of $AgCl_xBr_{1-x}$ alloys recently appeared using density-functional perturbation theory and employing the virtual crystal approximation or by means of the full potential linearized augmented plane wave method. Here, we suggest a simple theoretical model that enables the estimation of the isothermal compressibility of these alloys in terms of the elastic data of end members alone. The calculated values are in satisfactory agreement with the experimental ones. The present model makes use of an early suggestion that interconnects the Gibbs energy for the formation and/or migration of defects in solids with bulk properties.

# 1. Introduction

The silver halides exhibit interesting properties compared to the alkali halides, such as lower melting point and higher ionic conductivity. Silver halides are of great importance as photographic materials, as solid electrolytes and as liquid semiconductors (e.g., see Refs [1-5]). Although they all have the same NaCl structure, as the alkali halides, we emphasize that the elastic properties of the silver halides *cannot* be explained with the simple theories that successfully describe the elastic properties of the alkali halides [6].

Many experimental [6–12] and theoretical [13] studies have been carried out to understand the structural and the elastic properties, the phase transformation at high pressure, and the lattice dynamics of the AgBr, the AgCl and the AgBr$_{1-x}$Cl$_x$ ternary alloys. For example, recently, Shigeki Endou et al. [1] have measured the temperature dependence of the elastic constants in the silver halide crystals, above room temperature, by using the Resonant Ultrasound Spectroscopy method [14]. As a second example, we refer to Ref. [15], in which the elastic properties and the lattice dynamics of AgBr$_{1-x}$Cl$_x$ have been studied as a function of the composition (x) in the NaCl (B1) phase, by using the density-functional perturbation theory and employing the virtual-crystal approximation. Thirdly, Amrani et al. [2], in order to help understand and control the alloy system between AgCl and AgBr and behavior of bowing and related properties, have investigated the effect of the Cl concentration on the structural and electronic properties of the AgCl$_x$Br$_{1-x}$ alloys, with Cl contents between 0 and 1, using the Full Potential-Linearized Augmented Plane Wave method. The effect of composition on bulk modulus was investigated. This property was found to depend nonlinearly on alloy composition x.

The question arises whether one can determine the values of bulk modulus of a $AgCl_xBr_{1-x}$ mixed system, solely in terms of the elastic data of the end members AgBr and AgCl. This paper aims to answer this question. We employ here a simple model, that has been also recently [16] used for the calculation of the compressibility of multiphased mixed alkali halides crystals grown by the melt method [17] using the miscible alkali halides, i.e., NaBr and KCl, which have a simple cubic space lattice of the NaCl-type and measured in a detailed experimental study by Padma and Mahadevan [17]. This model has been also successfully applied [18] to the mixed crystal $NH_4Cl_{1-x}Br_x$ considering that $NH_4Cl$ and $NH_4Br$ have a simple cubic space lattice structure of the CsCl-type. In this paper we report the remarkable finding that this simple model produces in the case of $AgCl_xBr_{1-x}$ alloys equally successful results as in the mixed alkali halides and mixed ammonium halides despite the aforementioned significant differences in their physical properties and especially the lack [6] of a unified explanation with simple theories of the elastic properties of silver halides and alkali halides, as mentioned above. We emphasize, however, that the procedure through which this simple model is applied here to $AgCl_xBr_{1-x}$ *differs essentially* from the one followed for its application to mixed alkali and ammonium halides as it is explained in the last paragraph of the next section.

## 2. The method

We first recapitulate the model that explains how the compressibility $\kappa(=1/B)$ of a mixed system $A_xB_{1-x}$ can be determined in terms of the compressibilities of the two end members A and B. Let us call the two end members $A$ and $B$ as pure components (1) and (2), respectively and label $\upsilon_1$ the volume per "molecule" of the pure component (1) (assumed to be the major component in the aforementioned mixed system $A_xB_{1-x}$), $\upsilon_2$ the volume per "molecule" of the pure component (2). Furthermore, let denote $V_1$ and $V_2$ the corresponding molar volumes, i.e. $V_1 = N\upsilon_1$ and $V_2 = N\upsilon_2$ (where $N$ stands for Avogadro's number) and assume that $\upsilon_1 < \upsilon_2$. Defining a "defect volume" [19] $\upsilon_{2,1}^d$ as the increase of the volume $V_1$, if one "molecule" of type (1) is replaced by one "molecule" of type (2), it is evident that the addition of one "molecule" of type (2) to a crystal containing $N$ "molecules" of type (1) will increase its volume by $\upsilon_{2,1}^d + \upsilon_1$ (see Chapter 12 of Ref. [19] as well as Ref. [20]). Assuming that $\upsilon_{2,1}^d$ is independent of composition, the volume $V_{N+n}$ of a crystal containing $N$ "molecules" of type (1) and $n$ "molecules" of type (2) can be written as:

$$V_{N+n} = N\upsilon_1 + n(\upsilon_{2,1}^d + \upsilon_1) \quad \text{or} \quad V_{N+n} = [1+(n/N)]V_1 + n\upsilon_{2,1}^d \qquad (1)$$

The compressibility $\kappa$ of the mixed crystal can be found by differentiating Eq.(1) with respect to pressure which gives:

$$\kappa V_{N+n} = [1+(n/N)]\kappa_1 V_1 + n\kappa_{2,1}^d \upsilon_{2,1}^d \quad \text{or}$$

$$\kappa V_{N+n} = \kappa_1 V_1 + (n/N)\left[\kappa_{2,1}^d N \upsilon_{2,1}^d + \kappa_1 V_1\right] \tag{2}$$

where $\kappa_{2,1}^d$ denotes the compressibility of the volume $\upsilon_{2,1}^d$, defined as

$$\kappa_{2,1}^d \equiv -(1/\upsilon_{2,1}^d) \times (d\upsilon_{2,1}^d/dP)_T .$$

Within the approximation of the hard-spheres model, the "defect–volume" $\upsilon_{2,1}^d$ can be estimated from:

$$\upsilon_{2,1}^d = (V_2 - V_1)/N \quad \text{or} \quad \upsilon_{2,1}^d = \upsilon_2 - \upsilon_1 \tag{3}$$

Thus, since $V_{N+n}$ can be determined from Eq.(1) (upon considering Eq.(3) ), the compressibility $\kappa$ can be found from Eq.(2) if a procedure for the estimation of $\kappa_{2,1}^d$ will be employed. In this direction, we adopt a thermodynamical model for the formation and migration of the defects in solids described below which has been of value in various categories of solids including [21-26] metals, ionic crystals, rare gas solids etc as well as in high T$_c$ superconductors [27] and in complex ionic materials under uniaxial stress [28] that emit electric signals before fracture, in a similar fashion with the signals observed [29, 30] before the occurrence of major earthquakes.

According to the latter thermodynamical model, the defect Gibbs energy $g^i$ is interconnected with the bulk properties of the solid through the relation $g^i = c^i B\Omega$ (usually called $cB\Omega$ model) where $B$ stands for the isothermal bulk modulus ($=1/\kappa$), $\Omega$ the mean volume per atom and $c^i$ is dimensionless quantity. (The superscript $i$ refers to the defect process under consideration, e.g. defect formation, defect migration and self-diffusion activation). By differentiating this relation in respect to pressure $P$, we find that defect volume $v^i$ $[=(dg^i/dP)_T]$. The compressibility $\kappa^{d,i}$ defined by $\kappa^{d,i}$ $[\equiv -(d\ell n v^i/dP)_T]$, is given by [22, 23]:

$$\kappa^{d,i} = (1/B) - (d^2B/dP^2)/[(dB/dP)_T - 1] \qquad (4)$$

This relation states that the compressibility $\kappa^{d,i}$ does *not* depend on the type i of the defect process. Thus, it is reasonable to assume now that the validity of Eq. (4) holds also for the compressibility $\kappa^d_{2,1}$ involved in Eq. (2), i.e.,

$$\kappa^d_{2,1} = \kappa_1 - (d^2B_1/dP^2)/[(dB_1/dP)_T - 1] \qquad (5)$$

where the subscript "1" in the quantities at the right side denotes that they refer to the pure component (1). The quantities $dB_1/dP$ and $d^2B_1/dP^2$, when they are not experimentally accessible, can be estimated from the modified Born model according to [19, 20]:

$$dB_1/dP = (n^B + 7)/3 \text{ and } B_1(d^2B_1/dP^2) = -(4/9)(n^B + 3) \qquad (6)$$

where $n^B$ is the usual Born exponent. This is the procedure that has been successfully applied in Ref. [16] for the multiphased mixed alkali crystals, as well as in mixed ammonium halides [18]. Attention is drawn, however, to cases like $AgCl_xBr_{1-x}$ where the Born model does not provide an adequate description [6], as does for alkali halides.

Thus, here, for the case of $AgCl_xBr_{1-x}$ we shall solely rely on Eq. (4), but not on Eq. (6). In other words in our former publications [16, 18] dealt either with mixed alkali halides or with ammonium halides, we calculated the first and second pressure derivatives of the bulk modulus on the basis of Eq. (6) –obtained from the modified Born model- and then inserted them into Eq. (4). On the other hand in the present case of $AgCl_xBr_{1-x}$ we do not use at all the modified Born model, but we insert into Eq. (4) the first and second pressure derivative of the bulk modulus deduced from the elastic data of AgBr under pressure using a least squares fit to a second order Murnaghan equation as it will be described in the next section.

**3. Results**

Let us apply this procedure to the mixed system: AgBr-AgCl. In this application we shall intentionally take as starting material AgBr (1) ($V_1$=28.996 cm$^3$/mole) and by considering that for the pure AgCl (2) the volume is $V_2$=25.731 cm$^3$/mole, one gets $N\upsilon^d = V_2 - V_1 = -3.265\,cm^3$. We now consider the adiabatic values measured for various compositions in Ref. [6] and transform them to the isothermal ones with the standard thermodynamical procedure described in Ref. [19]. Using these isothermal $\kappa$-values, for various compositions x, we actually find that $\kappa V_{N+n}$ versus $n/N$ is a straight line the slope of which, according to the Eq. (2), is $\kappa_{2,1}^d\ \kappa_{2,1}^d(N\upsilon^d) + \kappa_1(AgBr)V_1(AgBr) = 63.99\text{x}10^{-2}$ cm$^3$GPa$^{-1}$ By inserting the $\upsilon^d$-value we find $\kappa_{2,1}^d = 3.947 \times 10^{-2}$ GPa$^{-1}$. Note that, the $\kappa_{2,1}^d$-value is appreciably higher than the compressibility of AgBr ($\kappa_1 = 2.645 \times 10^{-2}$ GPa$^{-1}$) and AgCl ($\kappa_2 = 2.398 \times 10^{-2}$ GPa$^{-1}$), as expected from thermodynamic arguments forwarded in Ref. [19]

We now proceed to the calculation of $\kappa_{2,1}^d$ on the basis of Eq. (5), by using the elastic data under pressure [31], which are well described if the expansion of the isothermal bulk modulus is carried out to second order, i.e.,

$$-\left(\frac{\partial P}{\partial \ln V}\right) = B(P) = B_0 + \left.\frac{dB_0}{dP}\right|_T P + \frac{1}{2}\left.\frac{d^2 B_0}{dP^2}\right|_T P^2$$

the investigation of which yields a second order Murnaghan equation (the subscript "0" corresponds to values close to zero pressure). The resulting expression for the bulk modulus of AgBr was found to be [31]: $B(P) = 377.7 + 7.49P - \frac{1}{2}(0.0287)P^2$ where B and P are in kilobars thus $\left.\frac{dB}{dP}\right|_T = 7.49$ and $\left.\frac{d^2 B}{dP^2}\right|_T = -0.0287$.

By inserting these values into Eq. (5) we find $\kappa_{2,1}^d = 7.06 \times 10^{-2}$ GPa$^{-1}$, with $\kappa_1 = 2.645 \times 10^{-2}$ GPa$^{-1}$ the isothermal compressibility for AgBr [18].

From Eq. (1) we calculate the volume $V_{N+n}$ for each concentration of the mixed crystals and from Eq. (2) the values of the isothermal bulk modulus. All the calculated values for the isothermal bulk modulus are depicted with asterisks in Fig. 1 where they are plotted versus the composition (x). In the same figure, we also insert with crosses the experimental values deduced from the adiabatic values measured in Ref. [6] and transformed to the isothermal ones by the standard thermodynamical procedure [19], as already mentioned.

We now turn to the values of the adiabatic bulk modulus. The theoretical values calculated in Refs. [2] and [15] are plotted in Fig. 1 with solid circles and open reverse triangles, respectively. We also insert with open squares the values calculated by the aforementioned simple thermodynamical model, where we followed the same procedure as above, but by considering the adiabatic values instead of the isothermal ones. In the same plot, we also show with solid triangles the experimental adiabatic values of the bulk modulus as reported in Ref. [6]. An inspection of these values reveals that there exists a disparity between the values calculated in Refs. [2] and [15]. Furthermore, we see that the values resulted from the simple model discussed here lie between those calculated in Refs. [2] and [15].

## 4. Conclusions

Here, we made use of the key-concept that the volume variation produced by the addition of a "foreign molecule" to a host crystal can be considered as a defect volume. Then the compressibility $\kappa_{2,1}^d$ of this defect volume was calculated on the basis of an early thermodynamical model which interconnects the defects Gibbs energy with bulk properties. This way enables the estimation of the isothermal compressibility of the rock-salt $AgCl_xBr_{1-x}$ alloys in terms of the elastic data of the pure constituents (i.e., AgBr and AgCl) alone. In all the composition range for which experimental data are available, the calculated values of the isothermal compressibility of these alloys are in reasonable agreement with the experimental ones. If we consider the adiabatic compressibility of these alloys, instead of the isothermal one, the values obtained by the present model lie between those resulted from the microscopic calculations carried out by other authors [2, 15].

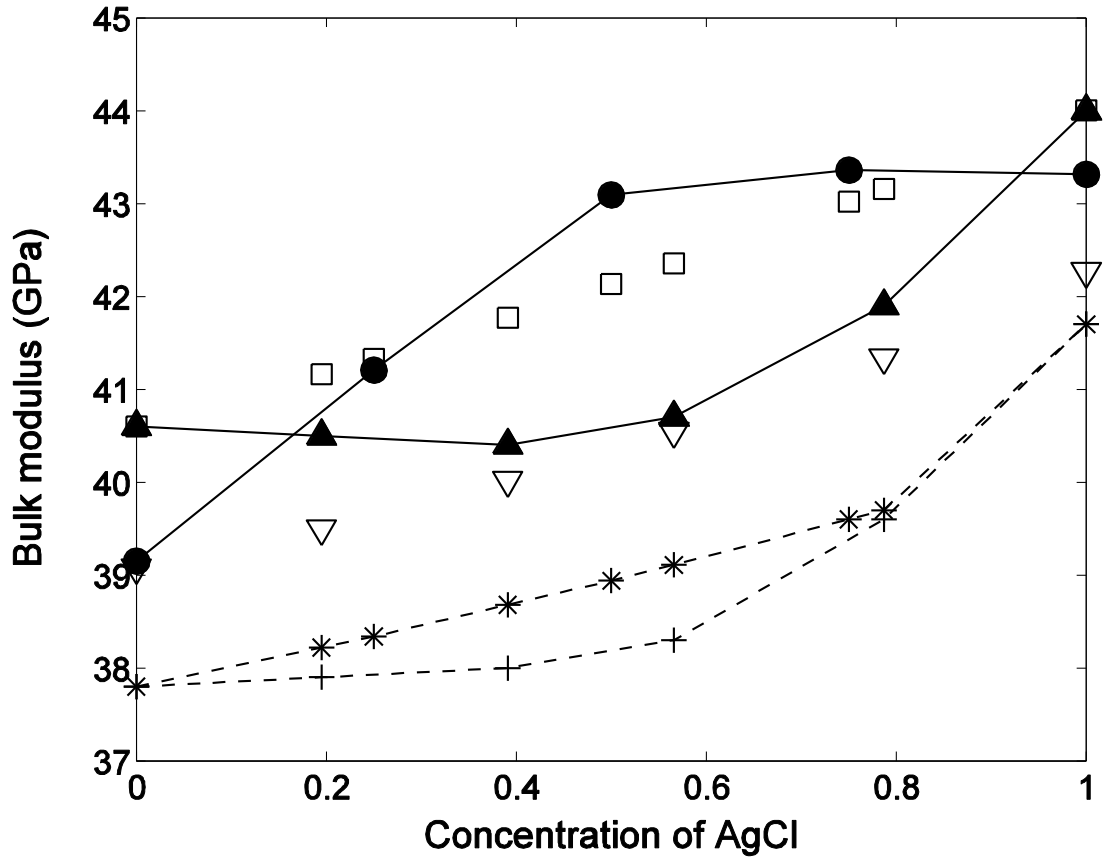

**FIG.1** The asterisks and the crosses mark the theoretical and the experimental values of the isothermal bulk modulus (broken lines). The latter come from the adiabatic values measured in Ref. [6] after transforming them to the isothermal ones by means of the standard thermodynamical manner (see Ref. [19]). We also plot for the adiabatic bulk modulus the theoretical (solid circles from Ref. [2], open reverse triangles form Ref. [15] and open squares from the simple model presented here) along with the experimental values [6] (solid triangles).